\documentclass[sigchi-a]{acmart}
\usepackage{booktabs} % For formal tables
\usepackage{ccicons}  % For Creative Commons citation icons

\usepackage{blindtext}
\usepackage{caption, subcaption}
\usepackage{breakurl}

\copyrightyear{2019} 
\acmYear{2019} 
\setcopyright{rightsretained}
\acmConference[CHI'19 Extended Abstracts] {CHI Conference on Human Factors in Computing Systems Extended Abstracts}{May 4--9, 2019}{Glasgow, Scotland UK}
\acmBooktitle{CHI Conference on Human Factors in Computing Systems Extended Abstracts (CHI'19 Extended Abstracts), May 4--9, 2019, Glasgow, Scotland UK}
\acmPrice{} 
\acmISBN{978-1-4503-5971-9/19/05}
\acmDOI{10.1145/3290607.3308461}
\begin{document}
\title{Automatic Speech Recognition Services: Deaf and Hard-of-Hearing Usability}

\author{Abraham Glasser}
\affiliation{%
  \institution{Rochester Institute of Technology}
  \city{Rochester}
  \state{NY}
  \postcode{14623}
  \country{USA} }
\email{abraham.glasser@gmail.com}

\renewcommand{\shortauthors}{F. Author et al.}

% REMOVED THIS TO GET MORE SPACE (ALSO REMOVED ACM REFERENCE FORMAT
% http://chi2019.acm.org/authors/chi-proceedings-format/ says we can do this.

% %
% % The code below should be generated by the tool at
% % http://dl.acm.org/ccs.cfm
% % Please copy and paste the code instead of the example below.
% %
% %
% % The code below should be generated by the tool at
% % http://dl.acm.org/ccs.cfm
% <concept_id>10003120.10003121.10011748</concept_id>
% <concept_desc>Human-centered computing~Empirical studies in HCI</concept_desc>
% <concept_id>10003120.10011738.10011776</concept_id>
% <concept_desc>Human-centered computing~Accessibility systems and tools</concept_desc>
% <concept_significance>300</concept_significance>
% <concept_id>10003120.10003123.10010860.10010858</concept_id>
% <concept_desc>Human-centered computing~User interface design</concept_desc>
% <concept_significance>300</concept_significance>

% \ccsdesc[500]{Human-centered computing~Empirical studies in HCI}
% \ccsdesc[300]{Human-centered computing~Accessibility systems and tools}
% \ccsdesc[500]{Human-centered computing~User interface design}

\begin{abstract}
Nowadays, speech is becoming a more common, if not standard, interface to technology. This can be seen in the trend of technology changes over the years. Increasingly, voice is used to control programs, appliances and personal devices within homes, cars, workplaces, and public spaces through smartphones and home assistant devices using Amazon's Alexa, Google's Assistant and Apple's Siri, and other proliferating technologies. However, most speech interfaces are not accessible for Deaf and Hard-of-Hearing (DHH) people. In this paper, performances of current Automatic Speech Recognition (ASR) with voices of DHH speakers are evaluated. ASR has improved over the years, and is able to reach Word Error Rates (WER) as low as 5-6\% \cite{google}\cite{ms}\cite{ibm}, with the help of cloud-computing and machine learning algorithms that take in custom vocabulary models. In this paper, a custom vocabulary model is used, and the significance of the improvement is evaluated when using DHH speech.
\end{abstract}


\maketitle

\section{Introduction and Previous Work}

The current and developing trend of speech interfaces can be seen in modern cars and home assistant devices. Such devices use Automatic Speech Recognition (ASR) to detect what is being spoken to them, and perform accordingly. Foreign accents and disfluencies in speech, historically, have had big impacts on the capability of ASR to understand human speech. However, current ASR technology is able to perform well, even if human speakers have these differences in speech. Current technology is able to do this with the help of cloud computing, machine learning, and sufficient datasets.

Glasser, Kushalnagar, and Kushalnagar did a preliminary study on using Deaf and Hard-of-Hearing (DHH) speech \cite{Glasser2017}. However, there have been significant advances in ASR since then. ASRs have been becoming more accurate, boasting Word Error Rates (WERs) as low as 5-6\%\cite{google}\cite{ms}\cite{ibm}. Also, there have been advancements where an user can provide custom language models for the ASR to have context awareness and improve accuracy.

In this study, a deeper analysis is done than the preliminary work done in \cite{Glasser2017}. Also, an analysis of WER improvement when using an ASR engine with custom language models is performed.

\begin{marginfigure}
    \frame{\includegraphics[width=\marginparwidth]{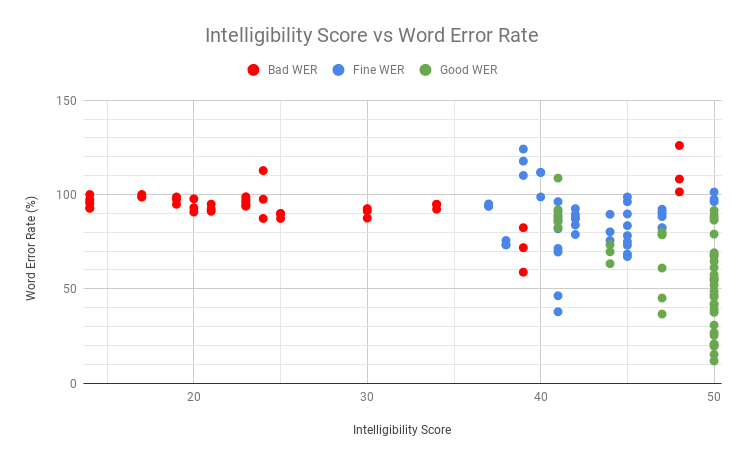}}
    \Description{Scatterplot of intelligbility scores}
    \caption{Scatterplot of intelligibility scores and WER for the audio database \\ ~ \\ }
    \label{fig:intelligibility_wer_scatter}
\end{marginfigure}

\begin{marginfigure}
    \includegraphics[width=\marginparwidth]{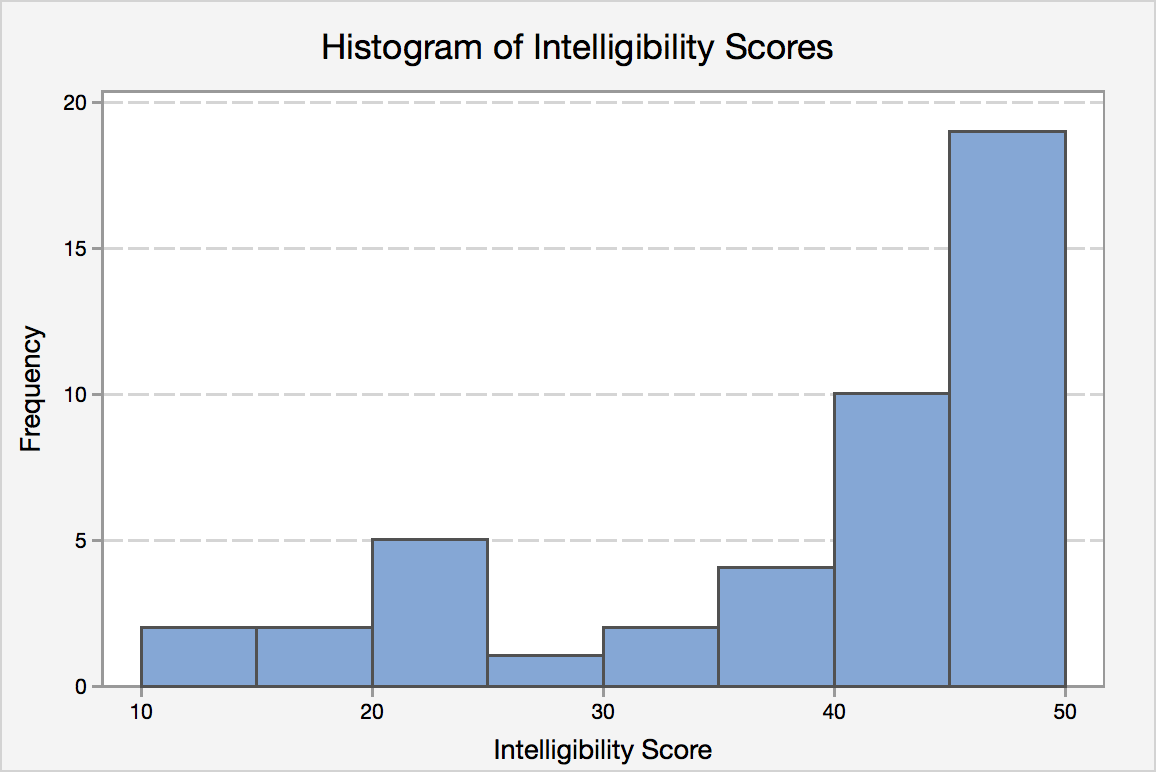}
    \Description{Histogram of intelligibility scores}
    \caption{Histogram showing frequency of intelligibility scores}
    \label{fig:intelligbility_histogram}
\end{marginfigure}

Even though ASR technology has improved dramatically over the past few years, and is now being incorporated in everyday technologies, it still has an usability challenge when it comes to the DHH population. DHH speech generally sounds different from hearing speech, and varies greatly between DHH individuals. DHH speech is often so variable that there is difficulty in understanding, even among experienced and inexperienced human listeners \cite{McGarr1983}.

\section{Methodology}

\subsection{Audio Dataset}

The dataset used in this study is a subset of a large speech dataset of 650 Deaf and Hard-of-Hearing (DHH) individuals at the National Technical Institute for the Deaf at Rochester Institute Technology, which has an enrollment of around 1100 DHH students. The dataset consists of DHH individuals who took the Clarke Sentences intelligibility test \cite{ClarkeSentences}. The test has 60 sentence lists, with 10 sentences per list. The sentences each have 10 syllables. The number of actual words varies across the sentences and lists. Each audio file has one DHH speaker reading one sentence list. The audio recordings were rated by a speech pathologist, who gave them an intelligibility score of 0 to 50. A score of 50 indicates that the speech is understood by the pathologist, while a score of 30 means difficult to understand, and a score of 0 means completely unintelligible.

In \cite{Glasser2017}, 45 audio files were chosen by a naive listener. 15 samples were rated "good", 16 samples were "fine" and 14 samples were "bad". These were determined by the naive listener who categorized the audio files in these three categories. The terms "good", "fine", and "bad" are used in this paper to refer to these categories. The average intelligibility score for the audio files in the "bad" category was 25, 43 for the "fine", and 48 for the "good".

The Waveform Audio File Format (filename extension .wav) container format was used, and the audio itself was encoded using PCM 16-bit little-endian encoding. This is high fidelity, uncompressed digital audio.

\vspace{-.2cm}
\subsection{ASR engines}

\begin{sidebar}
  ~\\
  
  \underline{MS} \\ \url{https://translator.microsoft.com/} \\
  
  \underline{MSPPT} \\ \url{https://translator.microsoft.com/help/presentation-translator/} \\
  
  \underline{IBM} \\ \url{https://speech-to-text-demo.ng.bluemix.net/} \\

  \caption{Links to the ASRs services used in this study}
  \label{bar:sidebar_links}
\end{sidebar}

For this study, the modern and widely used Automatic Speech Recognition (ASR) engines are the best fit, since they are generally being incorporated in everyday technology, and are also freely available for public use. Of these, the Microsoft Translator Speech API and the IBM Watson Speech to Text service were used. The web version of Microsoft Translator was used for the "base" model of the ASR engine, while the Presentation Translator for Microsoft PowerPoint plug-in was used for the "custom" model of the ASR engine.

For MSPPT, the entire list of Clarke Sentences was used as the keywords for the ASR engine to learn from. In theory, this should not decrease the accuracy of the MSPPT ASR compared to MS. MS and MSPPT are essentially the same ASR, with the exception that MSPPT has some "training" from vocabulary keywords that are given, and learns from it to improve accuracy. Throughout this paper, "customization" refers to this Context awareness.

IBM Watson Speech to Text was selected because it is available as a demo in their website, easily accessible, free to use, and was developed by a well-known large corporation.

Links to the ASRs used are provided in Sidebar \autoref{bar:sidebar_links}. They are all available for public use. They are also commercially available and continue to be improved on by the corporations.

\vspace{-.2cm}
\subsection{WER Analysis}

Word Error Rate (WER) is a standard measure of how accurate an ASR engine is. In this study, the National Institute of Standards and Technology Speech Recognition Scoring Toolkit (SCTK) Version 2.4.0.4 was used. This is freely available for people to use \cite{sctk}. The SCTK compares the reference ("truth script") and the hypothesis (output from ASR engine) transcripts, and calculates the WER. The transcripts are aligned, and the number of word substitutions, deletions, and insertions are found. The total of these divided by the total number of words in the reference transcript is the WER.

During the process of WER analysis, both the hypothesis and reference transcripts were converted to lowercase. Also, the newlines were stripped, and the entire transcript is on one line with a new line at the end. All punctuation marks except for apostrophes were removed. All this was manually done before WER analysis to eliminate things that may be different but should not be penalized across the reference and hypothesis transcripts.

Also, in the recordings themselves, there are sentences that were spoken by the person recording the session. These sentences consisted of the date, and prompts for the number of the sentence in the list. For example: "Today is 1/5/2009. Reading Clarke Sentences list number 43. Number one.", "Number two," and such. These were removed from the transcripts, and the reference transcripts did not contain these prompts. These "cleanings" of the transcripts were done to ensure that we did not account for information that was spoken by the non-DHH individual.

\vspace{-.2cm}
\section{Results}

%%% BONFERRONI RESULTS
\begin{margintable}
    \vspace{-2cm}
    \caption{Bonferroni correlation results for each ASR engine's performance across each audio category. \\ ~ \\ T.P = 2-sample T-test P-values \\ ~ \\ B.P = Bonferroni Correlation P-values \\ (* means significant)}
    \label{tab:bonferroni}
    \begin{tabular}{r r c}
      ASR & T.P & B.P \\
      \toprule
      \multicolumn{1}{c}{\textbf{IBM}} &  &  \\
      Bad-Fine & .1927 & 1 \\
      Bad-Good & .0005 & .0045 * \\
      Fine-Good & .0045 & .0405 * \\
      \midrule
      \multicolumn{1}{c}{\textbf{MS}} &  &  \\
      Bad-Fine & .161 & 1 \\
      Bad-Good & .0001 & .0009 * \\
      Fine-Good & .0005 & .0045 * \\
      \midrule
      \multicolumn{1}{c}{\textbf{MSPPT}} &  &  \\
      Bad-Fine & .0629 & .5661 \\
      Bad-Good & .0001 & .0009 * \\
      Fine-Good & .0005 & .0045 * \\
      \bottomrule
      ~\\~\\
    \end{tabular}
\end{margintable}

\begin{marginfigure}
    \includegraphics[width=1\marginparwidth]{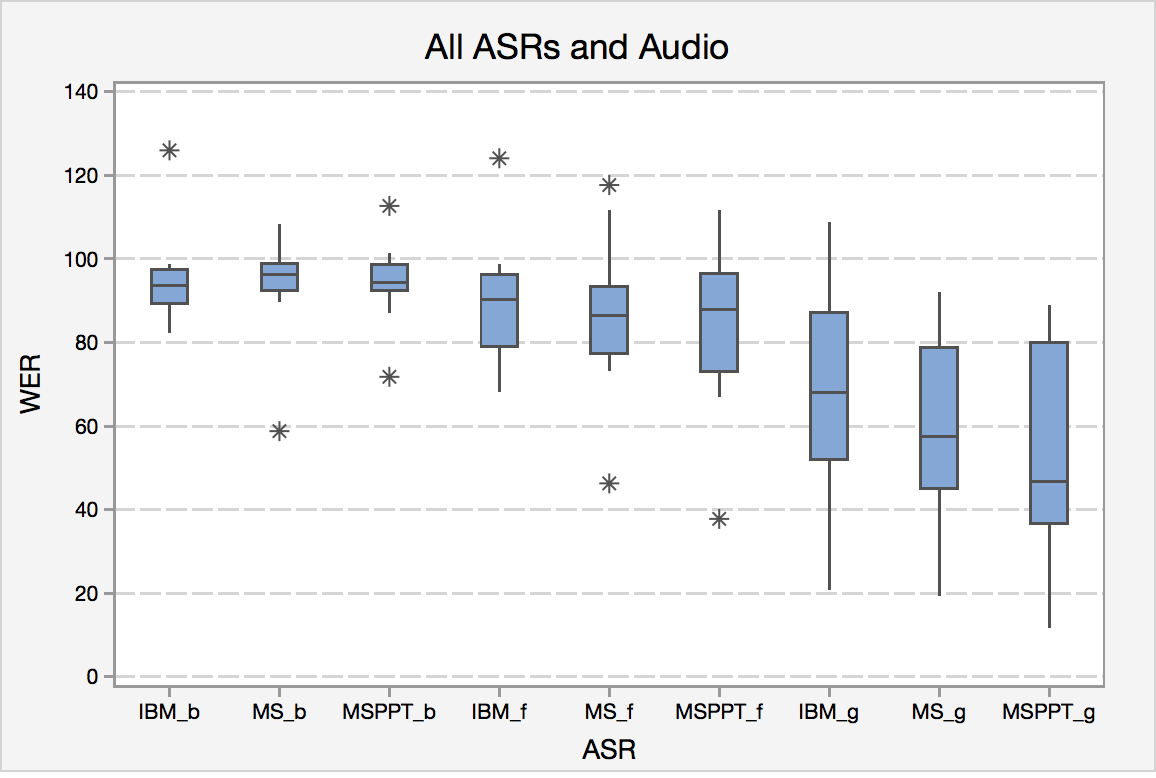}
    \Description{Side by side boxplots}
    \caption{Side by side boxplots of WER for all ASRs and audio categories \\ Asterisks (*) denote outliers}
    \label{fig:all_boxplots}
\end{marginfigure}

Three different ASRs ("IBM", "MS", and "MSPPT") and three different audio categories are used ("bad", "fine", and "good"). See the respective sections for in-depth explanations.

\vspace{-.1cm}
\subsection{WER for Deaf and Hard-of-Hearing Speech}

\autoref{fig:all_boxplots} shows a side by side comparison of all the boxplots for each ASR engine in each audio category. IBM\_b refers to the "IBM" ASR engine in the "bad" audio category, MS\_f refers to the "MS" ASR engine in the "fine" audio category, and so forth. As seen in these boxplots, all of the ASR engines in the "bad" and "fine" audio category had a high WER and a small variance. The "good" audio category had a large variance but an overall lower WER. This shows that for all the "bad" and "fine" audio, all the ASRs resulted in a very high WER, and a lower WER for the "good" audio, although not consistent. With voices of the non-DHH population, ASRs have improved drastically over the years and resulted in WERs as low as 5-6\% \cite{google}\cite{ms}\cite{ibm}.

%With voices of the DHH population, 
Even if a DHH voice is clear and sounds "good" to a naive listener, ASRs still do not always perform as expected. It was expected for the "good" audio to have a low WER but the variance in the data was very high, and the median of the WERs was above 45\% for all the ASRs in that audio category.

A one-sample t-test was performed for the WERs for each audio category. The 95\% confidence interval was found to be (91.338, 97.443) for "bad", (82.109, 91.316) for "fine", and (51.288, 66.068) for the "good" audio category. This shows that, with these ASRs, DHH speech will most likely result in a very high WER whether the audio sounds "bad" or "fine" to a naive listener and/or was assigned a intermediate intelligibility score by a speech pathologist.

\vspace{-.1cm}
\subsection{Improvements in WER for Deaf and Hard-of-Hearing Speech with Context Awareness}

Modern ASRs have reached low WERs with the help of customization, where they are provided a vocabulary/keyword list so they have Context awareness \cite{Dahl2012}. Context awareness helps the ASR engine better predict and identify what was said, and generally improves the accuracy of ASRs significantly.

\begin{figure}[h]
    \centering

\begin{subfigure}{0.32\textwidth}
  \includegraphics[width=\linewidth]{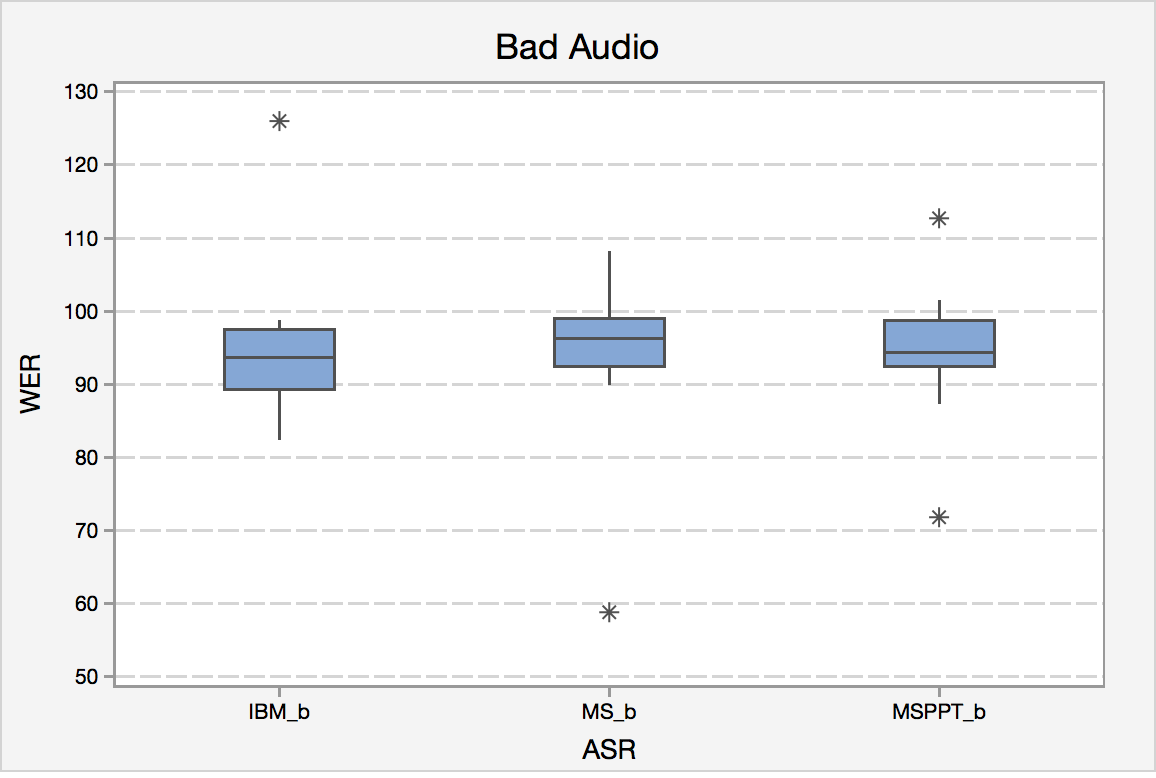}\Description{Boxplots for ASR engines for bad audio} 
  \caption{ASR engine boxplots for bad audio}
  \label{fig:bad_audio}
\end{subfigure}\hfil
\begin{subfigure}{0.32\textwidth}
  \includegraphics[width=\linewidth]{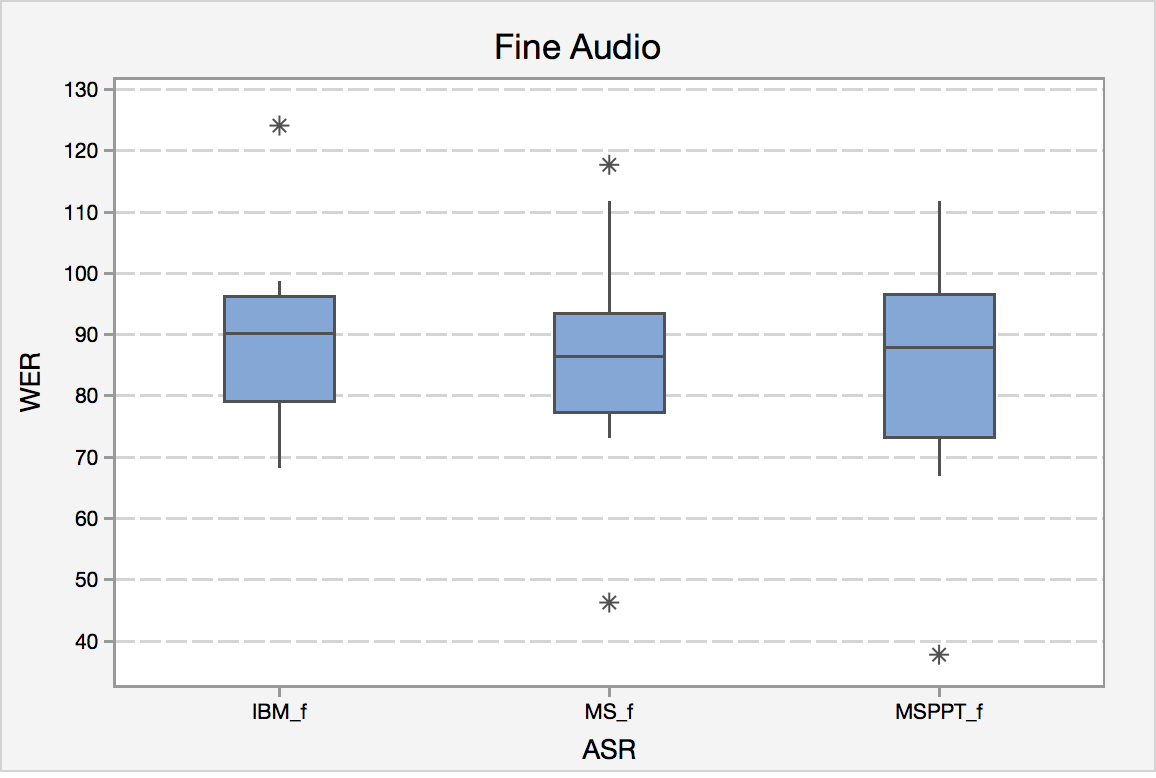}\Description{Boxplots for ASR engines for fine audio} 
  \caption{ASR engine boxplots for fine audio}
  \label{fig:fine_audio}
\end{subfigure}\hfil
\begin{subfigure}{0.32\textwidth}
  \includegraphics[width=\linewidth]{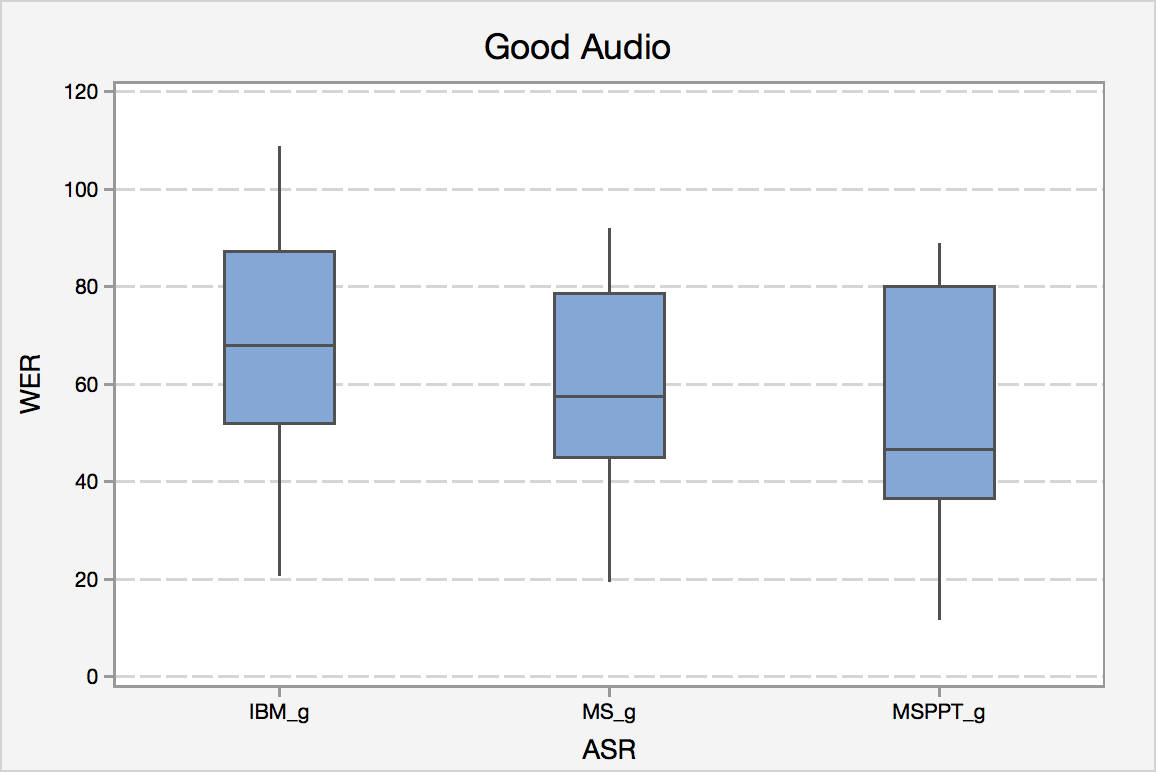}\Description{Boxplots for ASR engines for good audio} 
  \caption{ASR engine boxplots for good audio}
  \label{fig:good_audio}
\end{subfigure}
\medskip
\begin{subfigure}{0.32\textwidth}
  \includegraphics[width=\linewidth]{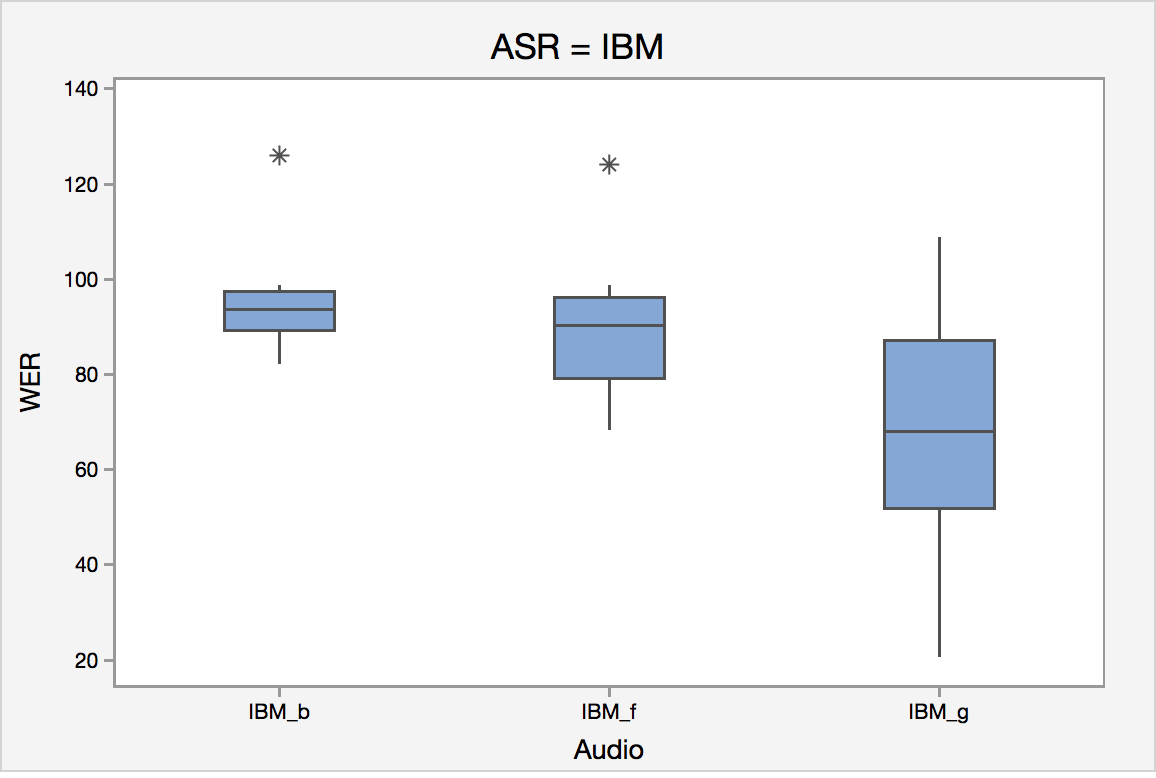}\Description{Boxplots for IBM ASR engine for all audio} 
  \caption{IBM ASR engine boxplots for all audio}
  \label{fig:ibm}
\end{subfigure}\hfil
\begin{subfigure}{0.32\textwidth}
  \includegraphics[width=\linewidth]{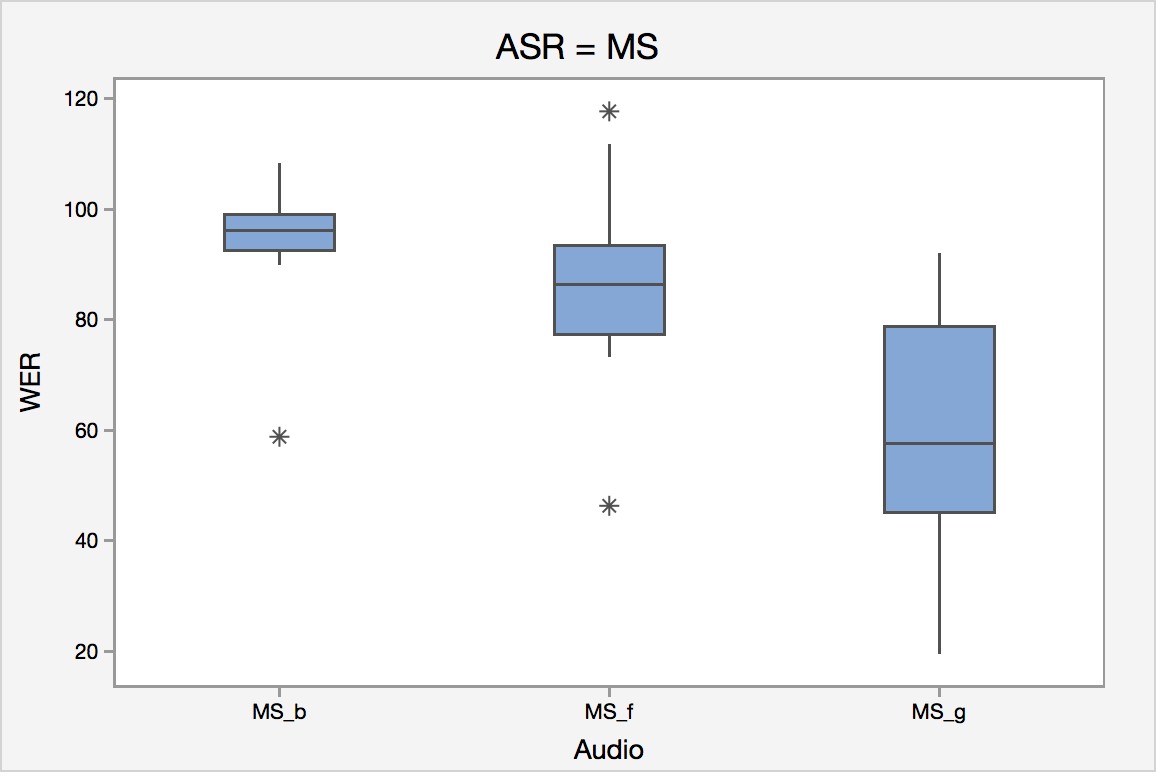}\Description{Boxplots for MS ASR engine for all audio} 
  \caption{MS ASR engine boxplots for all audio}
  \label{fig:ms}
\end{subfigure}\hfil
\begin{subfigure}{0.32\textwidth}
  \includegraphics[width=\linewidth]{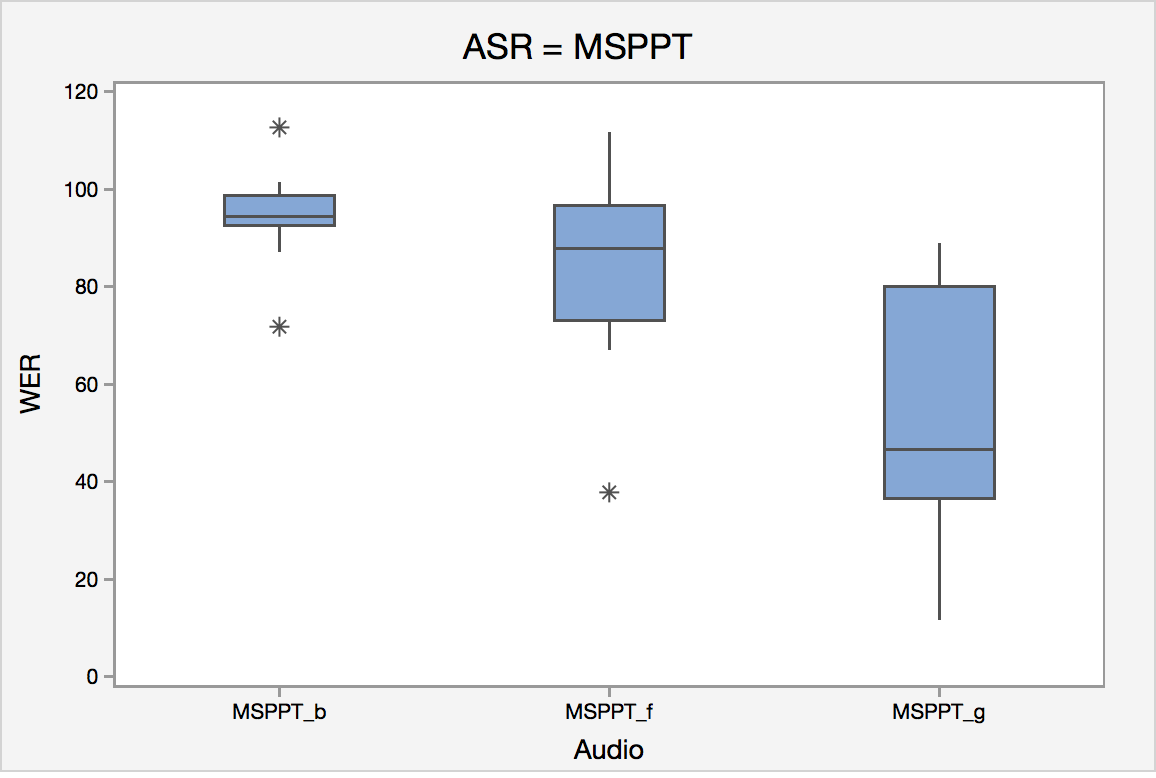}\Description{Boxplots for MSPPT ASR engine for all audio} 
  \caption{MSPPT ASR engine boxplots for all audio}
  \label{fig:msppt}
\end{subfigure}

\caption{Comparison of WER results boxplots for each ASR service and each audio category \\ Asterisks (*) denote outliers in these boxplots}
\label{fig:6_boxplots}
\end{figure}

However, there was not a significant improvement between the MS and MSPPT results. Even though the median WER for the "good" audio category improved by a little more than 10\%, the standard deviation was more than 20\%, since the results were very various. 2-sample T-tests were performed for all the MS vs. MSPPT results in all the audio categories, and the lowest P-value was .5472 from the "good" audio category. The results from the "bad" and "fine" showed there was basically no improvement between MS and MSPPT, as visualized in \autoref{fig:6_boxplots}.

%%%%%%%%%% 2 rows 3 columns figure with 6 boxplots
%%% from https://tex.stackexchange.com/questions/316435/how-to-put-6-images-in-3-columns-2-rows
%%%
\vspace{-.2cm}

\vspace{-.5cm}
\section{Conclusion}

The results for the 2-sample t-tests between the same ASR engine's performance in two different audio categories are shown in \autoref{tab:bonferroni}. A Bonferroni correlation test was done with these results, and all of the "bad" vs "fine" results were not significant. This verifies that the WER performance was not significantly different for any ASR engine between the "bad" and "fine" audio.

What this means is that DHH individuals would not be able to achieve equal WERs as the non-DHH population if their voice fell within the "bad" or "fine" audio categories. Even if their voice is "good", is still likely that they will get unpredictable results from the ASRs, as patterns in speech are very various within the DHH population, as pointed out by \cite{deafiot}.

As seen in \autoref{fig:6_boxplots}, the WER for the audio did not vary much between the ASR engines for the "bad" and "fine" categories, even when Context was provided for MSPPT. For the "good" audio category, where the DHH speech was rated very intelligible by a speech pathologist and put in the "good" category by a naive listener, the WER improved slightly between IBM and MS to MSPPT, albeit not significantly. The variance in the WER was much larger for all the ASR engines in the "good" audio category. This shows that you cannot yet use general DHH speech with ASRs.

\vspace{-.2cm}
\section{Future Work}

Our research shows that with enough data, it should be possible for ASRs to achieve consistent results with DHH speech, whether or not those results achieve low WERs. DHH speech is very various between individuals, and is even sometimes various within a specific DHH individual. If a DHH individual is able to achieve consistent results with an ASR, they might be able to use an acoustic model in addition to Context awareness to tailor the ASR to work with their speech.

ASR relies on having datasets to train with and learn from. Without enough data, ASR would not be able to achieve such low WERs as seen with voices of the non-DHH population. Companies and developers of modern ASRs have had access to large datasets of non-DHH speech, but not as much data for DHH speech. If a sufficiently large dataset of DHH speech is obtained and organized, then it is possible that ASRs will improve over time with this data and perform better than it has been with voices of DHH individuals.

\vspace{-.15cm}
\bibliography{bibliography}
\bibliographystyle{ACM-Reference-Format}

\end{document}